\documentstyle[12pt]{article}
 \def\be{\begin{equation}}
\def\ee{\end{equation}} 
\def\bea{\begin{eqnarray}}
\def\eea{\end{eqnarray}}

\def\delka{(1-n+ 2n\kappa)}
\def\deka{(-1+ 4\kappa)}

\def\Dsl{D \!\!\!\! /}
\def\G{{\bf \Gamma}}

\newcommand{\Section}[1]{\section{#1}\setcounter{equation}{0}}

\def\e{{\rm e}}

\def\nn{\nonumber}

\def\o{\over}

\def\b{\bar}
\def\l{\label}

\def\haf{\frac{1}{2}}
\def\){\right)}
\def\({\left(}

\begin{document}

\begin{flushright}
IPM/P-98/14  \\
hep-th/9805162
\end{flushright}

\pagestyle{plain}
\vskip .05in
\begin{center}

\Large{\bf Interacting Spinors-Scalars and AdS/CFT Correspondence}
\small
\vskip .15in

Amir M. Ghezelbash$_{a,b}$, Kamran Kaviani$_{a,b}$, Shahrokh Parvizi$_{b}$,
Amir H. Fatollahi$_{b,c}$

\vspace{.5 cm}
\small

{\it a)Department of Physics, Az-zahra University, 
P.O.Box 19834, Tehran, Iran}\\

\vspace{.3 cm}

{\it b)Institute for Studies in Theoretical Physics and Mathematics (IPM),}\\
{\it P.O.Box 19395-5531, Tehran, Iran}\\

\vspace{.3 cm}

{\it c)Department of Physics, Sharif University of Technology,\\
P.O.Box 11365-9161, Tehran, Iran}\\

\vspace{.3 cm}
{\sl E-mails: amasoud, kaviani, parvizi, fath@theory.ipm.ac.ir}

\begin{abstract}
By taking the interacting spinor-scalar theory on the $AdS_{d+1}$ space
we calculate the boundary CFT correlation functions using AdS/CFT
correspondence.

\end{abstract}
\end{center}
\Section{Introduction}
The correspondence between field theories in $(d+1)-$dimensional
Anti-de Sitter space and $d-$dimensional conformal field theories
has been studied in various aspects in the last few months. This
correspondence has been conjectured in the large $N$ limit of
superconformal gauge theories and the supergravity on $AdS_{d+1}$ spaces
\cite{1} and also has been studied in \cite{2} in connection with
the non-extremal black-hole physics. This suggested correspondence
has been made more precise in \cite{3,4,5}. The partition function
of any field theory on $AdS_{d+1}$ defined by,
\be \l{PAR}
Z_{AdS}[\phi _0]=\int _{{\phi _{o}}}{\cal D}\phi \e^{-S[\phi]},
\ee
where $\phi_0$ is the finite field defined on the boundary
of $AdS_{d+1}$ and the integration is over the field configurations
$\phi$ that go to the $\phi_0$ when one goes from the bulk of
$AdS_{d+1}$ to its boundary. According to the above mentioned works,
$Z_{AdS}$ is identified with the generating functional of the
boundary conformal field theory given by,
\be \l{PAR2}
Z_{CFT}[\phi_0]=<\e^{\int _{{M_{d}}}d^d x \sqrt g {\cal O}\phi_0}>,
\ee
for a quasi-primary conformal operator $\cal O$ on the $M_d$,
boundary of $AdS_{d+1}$. This correspondence were given in 
\cite{4} for a free massive scalar field and a
free $U(1)$ gauge theory. Some other examples as interacting massive scalar,
free massive spinor and massive interacting vector-spinor 
cases are studied in \cite{6}, \cite{7} and \cite{MV} respectively. 
 Also for classical gravity and type-IIB string theory
one can refer to \cite{Lu, Banks, Chal}. 
In all these cases, the exact partition function
(\ref{PAR}) is given by the exponential of the action evaluated for a
classical field configuration which solves the classical equations
of motion. The result of calculations shows that the evaluated
partition function is equal to the generating functional
(\ref{PAR2}) of some conformal field theory with a quasi-primary
operator with a certain conformal weight.

The purpose of this letter is to investigate the above correspondence 
in the case of interacting spinors-scalars theory. 
The natural motivation for this study comes from the importance of
non-linear SUSY theories living on the world-volumes of D-branes, namely
supersymmetric DBI actions. The tension of D-branes 
is involved by $\e^{-\phi}$ ($\phi$ dilaton field) 
and by noting different powers
of fermions in the action one can find vertices with different numbers 
of spinors and scalars. Even in the low energy limit which  
DBI action is approximated by SUSY gauge theories, the gauge coupling 
constant is involved by $\e^\phi$ and so the terms with quadratic in 
spinors and different number of scalars will appear. 

In particular, we calculate the boundary CFT correlation 
functions using AdS/CFT correspondence to be in agreement with CFT 
expectations. In the case of correlation functions with quadratic 
in spinors we found that they are generated only from a surface 
term introduced in \cite{7} to prevent the vanishing on-shell 
action. This observation is done also in \cite{MV} for the 
vector-spinor theory.

\Section{On-shell action}
Let us start with the following action for the interacting 
spinors-scalars theory which may be taken as a generalization of 
Fermi's four fermion theory
on the $AdS_{d+1}$,
\be \l{AC}
S=S_b+S_f+S_{int},
\ee
where
\bea \l{SB}
S_b&=&\int_{AdS}d^{d+1} x\, \sqrt G\frac{1}{2}\((\nabla \phi)^2+M_b^2\phi ^2\),
\\  \l{SF}
S_f&=&S_{f0}+S_{f1}\nn\\
&=&\int_{AdS}d^{d+1} x\, \sqrt G\(\b \psi (\Dsl-M_f)\psi \)
+\kappa \lim _{\epsilon \rightarrow 0} \int
_{M_d^\epsilon}d^d x \sqrt {G_\epsilon}\b \psi \psi,\\ \l{SINT}
S_{int}&=&\int_{AdS}d^{d+1} x\, \sqrt G \left( \sum _{n=0,m=0}^\infty
\lambda _{2n,m} (\b \psi \psi)^n\phi ^m \right),
\eea
with $\lambda_{2,0}=0$, $\kappa$ is a constant and $M_d^\epsilon$ is a
closed $d-$dimensional submanifold of $AdS_{d+1}$
which approaches the boundary manifold $M_d$ of $AdS_{d+1}$ as $\epsilon
\rightarrow 0$. 
The equations of motion in the bulk of the $AdS_{d+1}$ are given by,
\bea \l{EQ}
(\nabla ^2-M_b^2)\phi&=&\sum _{n=0,m=0}^\infty m\lambda _{2n,m} (\b \psi \psi)^{n}
\phi ^{m-1},\nn\\
(\Dsl-M_f)\psi&=&-\sum _{n=0,m=0}^\infty n\lambda _{2n,m} (\b \psi \psi)^{n-1}\psi
\phi ^m,\nn\\
\b \psi(-\Dsl-M_f)&=&-\sum _{n=0,m=0}^\infty n\lambda _{2n,m} (\b \psi
\psi)^{n-1}\b \psi\phi ^m.
\eea
The necessity of adding the $S_{f1}$ to $S_f$
in the interacting theory on $AdS_{d+1}$ is as follows \cite{7}: 
it can be shown easily that the quadratic spinor terms in action vanish 
with on-shell fields. So in calculating 
the partition function the correlation functions 
which are quadratic in spinors can not be generated.
As we show in the following, only surface term in the action 
is responsible to generate the correlators with quadratic in spinors. 
In fact the surface
term is sufficient to produce correlation functions quadratic in
spinors and at tree level no modification of the action is required.

We use the $AdS_{d+1}$ space which is represented in
$x=(x_0,\vec{x})$ coordinates by the domain $x_0>0$ and its metric
is given by,
\be \l{METRIC}
ds^2={1\o {{x_0}^2}}\left( (dx_0)^2+d\vec x\cdot d\vec x\right).
\ee
The boundary $M_d$ is the hypersurface $x_0=0$ plus a single point 
at $x_0=\infty$ and the metric on the $M_d$ is conformally flat 
$d\vec x\cdot d\vec x$. Also we have $ \sqrt{G}=x_0^{-d-1}$ and 
$\sqrt{G_\epsilon}=\epsilon^{-d}$. 
By introducing the Green's functions as 
\footnote{The explicit form of ${\cal{S}}$, $\bar{\cal{S}}$ and
$\cal{G}$ can be found in \cite{7, MV, 6}.}
\bea\l{GREEN}
(\nabla ^2-M_b^2){\cal{G}}(x,x')=\frac{\delta(x-x')}{\sqrt{G}},\\
(\Dsl-M_f){\cal{S}}(x,x')=\frac{\delta(x-x')}{\sqrt{G}},\\
\bar{\cal{S}}(x,x') (-\Dsl-M_f)=\frac{\delta(x-x')}{\sqrt{G}},
\eea
one can represent the solutions of equations of motion as \cite{6,7,MV}
\bea 
\l{BFC}
\phi(x)&=& \epsilon^{-d} \int_{M_d} d^dx' 
\frac{\partial {\cal{G}}}{\partial x_0}|_{x_0=\epsilon} 
\;\phi_\epsilon(\vec{x'})
\nn\\
&+&\int_{AdS}d^{d+1}x'\sqrt{G} {\cal{G}}(x,x')\sum _{n,m=0}^\infty 
m\lambda_{2n,m}(\b \psi (x')\psi (x'))^{n}\phi^{m-1}(x'),
\\
\l{FC}
\psi (x)&=& -\epsilon^{-d} \int_{M_d} d^dx'
\bar{\cal{S}}(x,\vec{x'})\psi_-(\vec{x'})
\nn\\
&-&\int_{AdS}d^{d+1}x'\sqrt{G} {\cal{S}}(x,x')\psi (x')
\sum _{n,m=0}^\infty 
n\lambda_{2n,m}(\b \psi (x')\psi (x'))^{n-1}\phi^m(x'),
\\ \l{FCC}
\b \psi (x)&=&-\epsilon^{-d}  
\int _{M_{d}} d^dx'\b\psi _+(\vec {x'})
{\cal{S}}(\vec{x'},x)
\nn\\
&-&\int _{AdS}d^{d+1}x'\sqrt{G} \b \psi(x') \bar{\cal{S}}(x',x)
\sum _{n,m=0}^\infty n\lambda_{2n,m} (\b \psi (x')\psi (x'))^{n-1}\phi^m(x'),
\eea
where $\b\psi_+(\vec x)$ and $\psi_-(\vec x)$ are the boundary value
of fields $\b\psi(x)$ and $\psi(x)$ 
with $\Gamma_0 \psi_-(\vec x)=-\psi_-(\vec x)$ and 
$\b\psi_+(\vec x) \Gamma_0=\b\psi_+(\vec x)$. 

By defining 
\bea\l{FREE1}
\phi^{(0)}(x)&\equiv& \epsilon^{-d} \int_{M_d} d^dx' 
\frac{\partial {\cal{G}}}{\partial x_0}|_{x_0=\epsilon} 
\;\phi_\epsilon(\vec{x'}),
\\ \l{FREE2}
\psi^{(0)} (x)&\equiv& -\epsilon^{-d} \int_{M_d} d^dx'
\bar{\cal{S}}(x,\vec{x'})\psi_-(\vec{x'}),
\\ \l{FREE3}
\b \psi^{(0)} (x)&\equiv&-\epsilon^{-d}  
\int _{M_{d}} d^dx'\b\psi _+(\vec {x'})
{\cal{S}}(\vec{x'},x),
\eea
one finds that $\phi^{(0)}$, $\psi^{(0)}$ and $\b \psi^{(0)}$ satisfy
the free equations of motion, so external legs in Feynman diagrams 
should be replaced by them. In a Born approximation 
one finds for the solutions of equations of motion as
\bea
\l{sol1}
\phi(x)&=& \phi^{(0)}(x)+\phi^{(1)}(x)+O(\lambda^2),
\\ \l{sol2}
\psi (x)&=& \psi^{(0)}(x)+\psi^{(1)}(x)+O(\lambda^2)
\\ \l{sol3}
\b \psi (x)&=&\bar\psi^{(0)}(x)+\bar\psi^{(1)}(x)+O(\lambda^2)
\eea
with
\bea
\phi^{(1)}(x)\equiv
\int_{AdS}d^{d+1}x'\sqrt{G} {\cal{G}}(x,x')\sum _{n,m=0}^\infty 
m\lambda_{2n,m}(\b \psi^{(0)} (x')\psi^{(0)} (x'))^{n}
(\phi^{(0)}(x'))^{m-1},
\nn\\
{}
\\
\psi^{(1)}(x)\equiv
-\int_{AdS}d^{d+1}x'\sqrt{G} {\cal{S}}(x,x')\psi^{(0)} (x')
\sum _{n,m=0}^\infty 
n\lambda_{2n,m}(\b \psi^{(0)} (x')\psi^{(0)} (x'))^{n-1}(\phi^{(0)}(x'))^m,
\nn\\
{}
\\
\bar\psi^{(1)}(x)\equiv
-\int _{AdS}d^{d+1}x'\sqrt{G} \b \psi^{(0)}(x') \bar{\cal{S}}(x',x)
\sum _{n,m=0}^\infty n\lambda_{2n,m} (\b \psi^{(0)} (x')
\psi^{(0)} (x'))^{n-1}(\phi^{(0)}(x'))^m.
\nn\\
\eea
By inserting the solutions (\ref{sol1}), (\ref{sol2}) and (\ref{sol3})
in the action and using the definitions (\ref{FREE1}), (\ref{FREE2}) 
and (\ref{FREE3}) and equations of motion one finds for the action
\footnote{ From now we ignore the pure scalar terms which have been 
studied in detail in \cite{6}.}
\bea \l{ACONSH}
S_{ON\;SHELL}&=&
\kappa \lim _{\epsilon \rightarrow 0} \int
_{M_d^\epsilon}d^d x \sqrt {G_\epsilon}\b \psi_+^{(0)} \psi_-^{(0)}
\nn\\
&+&  \int_{AdS}d^{d+1} x \sqrt {G}
\(\sum _{n=1,m=0}^\infty \delka
\lambda _{2n,m} (\b \psi^{(0)} \psi^{(0)})^n (\phi^{(0)}) ^m\)
\nn\\
&+& O(\lambda^2),
\eea
which shows that the on-shell action finds quadratic terms in 
spinors ($n=1$) only via the surface term in (\ref{AC}). It also 
shows that the other non-quadratic terms are multiplied by a 
 constant involving $\kappa$.

\Section{Correlation functions}

\subsection{$(2n,0)-$point functions (pure spinor)}
 
The boundary term (\ref{ACONSH}) gives the correlation
function of a pseudo-conformal operator ${\cal O}^\alpha $ and its
conjugate $\bar {\cal O}^\beta $ on the $M_d$ with the following
two-point function \cite{7},
\bea\l{TWO}
<\bar {\cal{O}}^\beta (\vec x){\cal{O}}^\alpha (\vec{x'})>=
\Omega^{\alpha\beta}
(\vec x,\vec {x'}),
\eea
where $\Omega (\vec{x},\vec {x'})=2\pi ^{d/2}\Gamma (M_f+1/2)/ \Gamma(M_f+
{{d+1}\o 2})
\mid \vec{x}-\vec{x'}\mid^{-d-1-2M_f}
(\vec x-\vec {x'})\cdot\G$.

Now, the interaction term in (\ref{ACONSH}) gives 
a contribution to the partition function as
\bea \l{ZI}
&~&Z_{AdS}[\b\psi_+, \psi_-]=
\exp\left(-\sum_{n=2}^\infty \delka \lambda_{(2n,0)}\int
d^{d+1}x\;x_0^{(n-1)(d+1)+2nM_f} 
\right.
\nn\\
&~&\left.
\prod_{i=1}^{2n}
d^dy_i\frac{1}{(x_0^2+|x-y_i|^2)^{\frac{d+1}{2}+M_f}}  
\prod_{j=1}^n
\b\psi_+(y_{2j-1})
(\vec y_{2j-1}-\vec y_{2j})\cdot \G\psi_-(y_{2j}) \right),
\eea
from which one can find the $(2n,0)-$point function,
\bea\l{TWON}
&~&G_{(2n,0)}(y_1,\cdots ,y_{2n})_{\mu_1,\cdots ,\mu_{2n}}=
\delka \lambda_{(2n,0)}\int d^{d+1}x\;x_0^{(n-1)(d+1)+2nM_f}
\nn\\
&~&\prod_{i=1}^{2n}
\frac{1}{(x_0^2+|x-y_i|^2)^{\frac{d+1}{2}+M_f}}
\times\prod_{j=1}^n
((\vec{y}_{2j-1}-\vec{y}_{2j})\cdot\G)_{\mu_{2j-1}\mu_{2j}}
\eea
By using the Feynman parameter technics one can find \cite{6}
\bea\label{Fey}
&~&\int d^{d+1}x
\frac{x_0^{-(d+1)+\sum_{i=1}^N \Delta_i}}
{\prod_{i=1}^N (x_0^2+\mid x-y_i\mid^2)^{\Delta_i}}=
\frac{\pi^\frac{d}{2} \Gamma(\frac{\sum \Delta_i}{2}-\frac{d}{2})
\Gamma( \frac{ \sum \Delta_i }{2}) }
{2\prod_i\Gamma(\Delta_i)}
\nn\\
&\times& \int_0^\infty d\alpha_1\cdot\cdot\cdot
d\alpha_n \delta(\sum\alpha_i-1)
\frac{\prod \alpha_i^{\Delta_i-1}}
{(\sum_{i<j} \alpha_i\alpha_j
y_{ij}^2)^\frac{\sum \Delta_i}{2}}.
\eea
Putting $\Delta_i=\Delta'\equiv\frac{d+1}{2}+M_f$ for all $i$
one may change the $(2n,0)$-point function
(\ref{TWON}) to a similar form which is suitable to investigate the conformal
properties.

For $n=2$, i.e. the $(4,0)-$point function we have
\bea    \l{PO}
&~&G_{(4,0)}(y_1, y_2,y_3,y_4)_{\alpha ,\beta ,\gamma ,\delta}
= \deka \lambda_{(4,0)}\pi^\frac{d}{2}
\frac{\Gamma(2\Delta'-\frac{d}{2})\Gamma(2\Delta')}{2\Gamma^4(\Delta')}
\int \prod_{i=1}^4 d\alpha_i
\delta(\sum_{i=1}^4\alpha_i-1)
\nn\\
&~&\times
\frac{(\alpha_1\alpha_2\alpha_3\alpha_4)^{\Delta'-1}
(\vec y_{12}\cdot\G)_{\alpha\beta}(\vec y_{34}\cdot\G)_{\gamma\delta}}
{(\alpha_1\alpha_2y_{12}^2+\alpha_1\alpha_3y_{13}^2+
\alpha_1\alpha_4y_{14}^2+\alpha_2\alpha_3y_{23}^2+
\alpha_2\alpha_4y_{24}^2+\alpha_3\alpha_4y_{34}^2)^{2\Delta'}},
\eea
where it can easily be shown that $G_{(4,0)}$ has all appropriate conformal
symmetries such as translation, scaling and inversion. To test
inversion, we can write (\ref{PO}) as follow,\cite{6}
\bea \l{POI}
&~&G_{(4,0)}(y_1, y_2,y_3,y_4)_{\alpha ,\beta ,\gamma ,\delta}=
\deka \lambda_{(4,0)}\nn\\&~&\frac{\Gamma(2\Delta'-\frac{d}{2})}{\Gamma(2\Delta')}
\frac{
(2\pi)^\frac{d}{2}
}
{(\eta \xi  \prod_{i<j} y_{ij})^{\frac{2}{3}\Delta'}}
(\vec y_{12}\cdot\G)_{\alpha\beta}
(\vec y_{34}\cdot\G)_{\gamma\delta}
\nn\\&~&
\times \int dz
F(\Delta', \Delta'; 2\Delta';1-\frac{(\eta+\xi)}{(\eta\xi)^2}-
\frac{4}{\eta\xi}\sinh^2 z) ,
\eea
where $\eta=\frac{y_{12}y_{34}}{y_{14}y_{23}}$ and
$\xi=\frac{y_{12}y_{34}}{y_{13}y_{24}}$.
Expanding the above relation in terms of $y_{ij}$ we have
\bea \l{POIU}
&~&G_{(4,0)}(y_1, y_2,y_3,y_4)_{\alpha ,\beta ,\gamma ,\delta}=
\deka \lambda_{(4,0)}\nn\\&~&\frac{\Gamma(2\Delta'-\frac{d}{2})}{\Gamma(2\Delta')}
\frac
{(2\pi)^\frac{d}{2}}
{(\eta \xi  \prod_{i<j} y_{ij})^{\frac{2}{3}\Delta_f}}
(\hat y_{12}\cdot\G)_{\alpha\beta}
(\hat y_{34}\cdot\G)_{\gamma\delta}
\nn\\
&~&\times \int dz
F(\Delta', \Delta'; 2\Delta';1-\frac{(\eta+\xi)}{(\eta\xi)^2}-
\frac{4}{\eta\xi}\sinh^2 z),
\eea
where $2\Delta_f\equiv\Delta'-1=d+2M_f$ and $\hat y_{ij}$
is the unit vector along $y_{ij}$,
which shows the invariance of $(4,0)-$point function
under inversion.
\subsection{$(2n,m)-$point functions (spinor-scalar)}
Here, we calculate the spinor-scalar
$(2n,m)-$point functions of boundary CFT. By using the relations
(\ref{SINT}), (\ref{FC}), (\ref{FCC}) and the following relation for the
bosonic field configuration in the bulk of $AdS_{d+1}$, \cite{6}
\be \l{BC}
\phi (x)=c\int _{M_{d}}d^dx'\phi _0(\vec x')\left({{x_0}\o
{{x_0}^2+\mid \vec x-\vec x'\mid ^2}}\right)^{\Delta _b}
\ee
where $\phi _0$ is the bosonic field $\phi (x)$ on the boundary and
$\Delta _b=d/2+\sqrt {M_b^2+{{d^2}\o 4}}$ and $c={{\Gamma (\Delta _b)}\o
{\pi ^{d/2}\Gamma (\Delta _b-d/2)}}$, one can calculate the partition function
to be
\bea \l{ZZ}
&~&Z_{AdS}[\psi_-,\b \psi _+,\phi]=
\exp\left(
-\sum_{n=1,m=1}^\infty
\delka\lambda_{(2n,m)}
\right.
\nn\\
&~&\left.
\int d^{d+1}x\; x_0^{-(d+1)+n+2n\Delta_f+m\Delta_b}
\int \prod_{k=1}^m d^dz_k
\frac{1}{(x_0^2+|x-z_k|^2)^{\Delta _b}} \phi_0(z_k)
\right.
\nn\\
&~&\left.
\times\int\prod_{i=1}^{2n}d^dy_i
\frac{1}{(x_0^2+|x-y_i|^2)^{\Delta_f+\haf}}
\prod_{j=1}^n
\b\psi_+(y_{2j-1})(\vec y_{2j-1}-\vec y_{2j})\cdot\G\psi_-(y_{2j})
\right),
\eea
from which one can find the $(2n,m)$-point functions
\bea
&~&G_{(2n,m)} (y_1,\cdots, y_{2n};z_1,\cdots,z_m)_{\mu_1,\cdots,\mu_{2n}}=
\delka\lambda_{(2n,m)}\nn\\&~&
\int d^{d+1}x\; x_0^{-(d+1)+n+2n\Delta_f+m\Delta_b}
\nn\\
&~& \prod_{k=1}^m
\frac{1}{(x_0^2+|x-z_k|^2)^{\Delta _b}}
\times\prod_{i=1}^{2n}
\frac{1}{(x_0^2+|x-y_i|^2)^{\Delta_f+\haf}}
\prod_{j=1}^n
((\vec y_{2j-1}-\vec y_{2j})\cdot\G)_{\mu_{2j-1},\mu_{2j}},
\nn\\
&~&
\eea
By using the relation (\ref{Fey}) and taking
\bea
\Delta_1=\ldots=\Delta_n=\Delta_f+\haf ,
\nn\\
\Delta_{n+1}=\ldots=\Delta_{N=n+m}=\Delta_b         ,
\eea
one  can obtain the explicit form of the $(2n,m)$-point functions.
For $n=1$ and $m=1$ one finds
\bea
&~&G_{2,1}(y_1,y_2;z)_{\gamma,\delta}=2\kappa\lambda_{2,1}
\frac{\pi^\frac{d}{2}
\Gamma(\frac{1+2\Delta_f+\Delta_b-d}{2})
\Gamma(\frac{1+2\Delta_f+\Delta_b}{2})}
{2\Gamma^2(\Delta_f+\haf)\Gamma(\Delta_b)}
\int_0^\infty d\alpha_1 d\alpha_2 d\alpha_3
\delta(\sum \alpha _i-1)
\nn\\
&~&\times\frac{ \alpha_1^{\Delta_f-\haf}\alpha_2^{\Delta_f-\haf}
\alpha_3^{\Delta_b-1}}
{ (\alpha_1\alpha_2(y_1-y_2)^2+
\alpha_1\alpha_3(y_1-z)^2+\alpha_2\alpha_3(y_2-z)^2)
^{(1+2\Delta_f+\Delta_b)/2}}
(\vec y_{12}\cdot\G)_{\gamma\delta}.
\eea
The integrals over $\alpha$'s can be done 
\bea
G_{2,1}(y_1,y_2;z)_{\gamma,\delta}&=&2\kappa\lambda_{2,1}
\frac{\pi^\frac{d}{2}
\Gamma(\frac{1+2\Delta_f+\Delta_b-d}{2})
\Gamma(\frac{1+2\Delta_f+\Delta_b}{2})}
{2\Gamma^2(\Delta_f+\haf)\Gamma(\Delta_b)}
\nn\\
&\times&
\frac{
B(\Delta_f+\haf,\frac{\Delta_b}{2})
B(\frac{\Delta_b}{2},\Delta_f+\frac{1-\Delta_b}{2})
}
{
(y_1-y_2)^{2\Delta_f-\Delta_b} 
(y_1-z)^{\Delta_b} 
(y_2-z)^{\Delta_b} 
}
(\hat y_{12}\cdot\G)_{\gamma\delta},
\eea
where $B$ is the beta function and $\hat y_{12}$ is the unit vector in 
the direction $(\vec y_1-\vec y_2)$.

The calculation of $(2,2)$ correlation function can be done as above.

\Section{Conclusion}
In this paper we apply the conjectured AdS/CFT correspondence to
spinor-scalar interacting field theory. In this way we calculated
the pure spinor and mixed spinor-scalar boundary CFT correlation 
functions. In all cases were done explicitly we found agreement with 
CFT expectations.

Our main result is related to the correlators with two spinor and 
any number of scalars which were introduced in the text as $(2,m)$-point 
functions. It is observed that $(2,m)$-point functions 
come only from a surface term introduced in \cite{7} 
(which is added to prevent the vanishing of the on-shell free action),
 and no more modification is required. 
 So we found that no modification is required
to get $(2,m)$ CFT correlators up to the first order of 
perturbation.

Also the effect of the surface term on the other correlators $(2n,m)$   
($n \geq 2$) found to be a multiplicative constant.

{\bf Acknowledgement}\\
We are grateful to K. Sfetsos for his helpful comment about the 
terms coming from the surface term and a missed factor two.


\end{document}